\documentclass[prb,twocolumn,showpacs,superscriptaddress,floatfix,amsfonts,amsmath]{revtex4}

\usepackage{amsmath}
\usepackage{amsfonts}

\usepackage{graphicx}
\usepackage{array}

\newcommand{\be}{\begin{equation}}
\newcommand{\ee}{\end{equation}}
\newcommand{\bea}{\begin{eqnarray}}
\newcommand{\eea}{\end{eqnarray}}
\newcommand{\beal}{\begin{align}}
\newcommand{\eeal}{\end{align}}
\newcommand{\bes}{\begin{equation} \begin{split}}
\newcommand{\ees}{\end{split} \end{equation}}
\newcommand{\down}{\downarrow}

\newcommand{\up}{\uparrow}
\newcommand{\f}{\frac}

\newcommand{\cobaltniobate}{\rm{CoNb}_2\rm{O}_6}

\newcommand{\deltaLL}{P_{r_{L}}^\leftarrow}
\newcommand{\deltaRR}{P_{r_{R}}^\rightarrow}
\newcommand{\order}[1]{{\cal O}(#1)}
\begin{document}

\title{Dynamical structure factor of magnetic Bloch oscillations at finite temperatures}
\author{Olav  F.~Sylju{\aa}sen}
\address{Department of Physics, University of Oslo, P.~O.~Box 1048 Blindern, N-0316 Oslo, Norway}

\date{\today}

\begin{abstract}
Domain-walls in one-dimensional Ising ferromagnets can undergo Bloch  oscillations when subjected to a skew magnetic field. Such oscillations imply finite temperature non-dispersive low-frequency peaks
in the dynamical structure factor which can be probed in neutron scattering. 
We study in detail the spectral weight of these peaks. Using an analytical approach based on an approximate treatment of a gas of spin-cluster excitations we give an explicit expression for the momentum- and temperature-dependence of the spectral weights. Generally the spectral weights increase with temperature $T$ and approaches the same order of magnitude as the spin-wave spectral weights at high temperatures.
We compare the analytical expression to numerical exact diagonalizations and find that it can, without any adjustable parameters, account for the $T$ and momentum-transfer dependence of the numerically obtained spectral weights in the parameter regime where the ratio of magnetic fields $h_x/h_z \ll 1$ and the temperature is $h_x < T <\sim  J_z/2$. 
We also carry out numerical calculations pertinent to the material $\cobaltniobate$, and find qualitatively similar results. 
\end{abstract}
\maketitle

\section{Introduction}
Bloch Oscillations (BO)\cite{Bloch,Zener}  is a pure quantum mechanical phenomenon where a particle in a periodic potential oscillates when acted upon by a constant external force.
Although BO are quite delicate, they have been observed in a range of systems in the last two decades\cite{BOsemiconductors,BOcondensates,BOoptics,BOultrasonics}. In addition BO have also been predicted\cite{KyriakidisLoss} to exist in one dimensional magnetic systems, so called magnetic Bloch oscillations (MBO). However, such MBO have so far not been observed experimentally. 

One way to search for MBO is to search for its energy spectrum which is the Wannier-Zeeman (WZ) ladder\cite{Wannier,Shiba} of equally spaced energy levels. Experimentally one can hope to probe the WZ ladder directly using neutron scattering. At low temperatures neutrons will induce transitions from the ground state to the WZ levels. However, the transition matrix elements of these processes are very small, thus the low temperature inelastic neutron scattering signal will be very weak\cite{Shinkevich1,Rutkevich}. In contrast, at temperatures comparable to the dominant magnetic scale the WZ levels will be thermally populated and neutrons can induce scattering between them. These transitions show up as peaks at low frequencies\cite{KyriakidisLoss} in the neutron scattering dynamical structure factor, and are the topic of this paper.

This paper focuses especially on the spectral weights of the finite temperature low-frequency peaks and how they depend on momentum transfer and temperature. These quantitative details are important in order to distinguish the peaks associated with BO from generic low-frequency peaks in other systems with Zeeman spin-split spectra. 

In order to calculate the spectral weights we use both an analytical and a numerical approach. The analytical approach is based on a gas of spin-cluster excitations with collisions treated in an approximate way. This goes beyond the one\cite{KyriakidisLoss} and two\cite{Shinkevich1} domain-wall approximations where collisions are neglected completely. The neglect of collisions can be expected to work well at low temperatures, but is more dubious at high temperatures where one expects frequent collisions between domain-walls. Also the one- and two domain-wall approximations restrict the state space severely to states that constitute the WZ levels. Since the existence of the low-frequency peaks is caused by the thermal population of the WZ levels it is important to verify that thermal population of other excluded states will not destroy them. For these reasons we validate our analytical result by carrying out numerical exact diagonalizations keeping all the states.        

While the analytical result presented in this paper is strictly valid for the Ising model in a skew magnetic field we expect it to hold qualitatively for a broader class of Hamiltonians. We substantiate that at the end of the paper by performing numerical exact diagonalizations of a Hamiltonian that also includes extra terms relevant for describing the material $\cobaltniobate$.  

The outline of the paper is the following. In chapter 2 we present the Hamiltonian and explain qualitatively why it should describe MBO. Chapters 3 and 4 explain the analytical calculation of the dynamical structure factor. The numerical exact diagonalization result is presented in chapter 5, while chapter 6 is devoted to a detailed comparison of the analytical and numerical results. In chapter 7 we present numerical results for the dynamical structure factor for parameter values relevant to the material $\cobaltniobate$, and chapter 8 concludes the paper.

\section{Hamiltonian}
The prediction that domain-walls in one-dimensional ferromagnets undergo BO in a magnetic field is rather general and holds for a range of Hamiltonians\cite{KyriakidisLoss}. In order to be definite we will focus on the one-dimensional spin-1/2 ferromagnetic Ising Hamiltonian with nearest-neighbor couplings in a uniform skew magnetic field
\be \label{Hamiltonian}
    H = - \sum_{i} \left( J_z S^z_i S^z_{i+1} +h_x S^x_i + h_z S^z_i \right). 
\ee
We will consider the case where the Ising coupling $J_z$ is the dominant one, and $J_z \gg h_z > h_x \geq 0$. Whenever numerical energies, frequencies or magnetic fields are quoted, they are in units of $J_z$. The lattice spacing and $\hbar$ are set to unity. 

The ground state of the Hamiltonian $(\ref{Hamiltonian})$ resembles closely the ferromagnetic all spins up state as quantum fluctuations are small. Excitations can be classified according to how many domain-walls they contain. A domain-wall is an antiferromagnetic arrangement of two neighboring spins and costs an energy $J_z/2$. It is these domain-walls that undergo BO in a skew magnetic field. To see why, consider first the transverse field $h_x$. It couples to the spin operator $S^x$ that flips a spin. When $J_z$ is large the main effect of the transverse field is to flip a spin adjacent to a domain-wall which causes it to move, and will broaden the domain-wall excitations into a band. Thus the role of $h_x$ is to mimic a periodic potential.  
The longitudinal magnetic field, $h_z>0$, plays the role of the external force as it will pull the domain-wall in the direction of the down-spins. 
The resulting BO of a single domain-wall in this system was recently studied in real-time numerical simulations\cite{Cai}.   

However, in contrast to other systems with BO, a single domain-wall cannot exist in isolation in the magnetic system. The competition between the longitudinal magnetic field and the Ising coupling means that the system cannot have arbitrarily long spin-down domains. Instead new domain-walls will be created, and the excitations will be pairs of domain-walls, i.e. domain-wall/(anti)-domain-wall bound states. Nevertheless, the magnetic field still causes the domain-walls to oscillate, thus MBO in domain-wall ferromagnets resemble breather oscillations.

\section{Spin-cluster excitations}
At $h_x=0$ the excitations are domains of $l$ consecutive down-spins surrounded by up-spins. For $h_x>0$ this is slightly altered, and the excitations will be superpositions of such domains with different numbers of down-spins\cite{Fogedby78}. These excitations have been termed multiple-magnon bound states\cite{TorranceTinkham}, but we will call them spin-cluster excitations for simplicity. A spin-cluster state can be described mathematically by  
labeling the state with $l$ consecutive down-spins starting at site $j$ as $|j,l\rangle$.
Defining Fourier-transformed states as $|p,l\rangle = \f{1}{\sqrt{N}} e^{-ipl/2} \sum_j e^{-ipj}|j,l\rangle$, and writing energy eigenstates as superpositions of these,
$|p,n\rangle = \sum_{l=1} \psi_{p,n}(l) |p,l\rangle$,  
the energy eigenvalue equation in the sector with two domain-walls takes the form
\begin{multline} \label{twodomainwalleq}
h_x \cos(p/2) \left[ \psi_{p,n}(l+1)+\psi_{p,n}(l-1) \right] = \\ 
\left( J_z + l h_z -E_{p,n} \right) \psi_{p,n} (l) \qquad ,l=1,2,\ldots      
\end{multline}
where we have defined $\psi_{p \neq 0,n}(0)=0$, and neglected all terms that change the number of domain-walls. This is expected to be a good approximation for large $J_z$.  

Eq.~(\ref{twodomainwalleq}) resembles the recursion relation for Bessel functions and has the solution
\be
   E_{p,n} = J_z +h_z \nu_{p,n}
\ee
and
\be
 \psi_{p,n}(l)= C_n J_{l-\nu_{p,n}}(z_p) \label{psi}
\ee
where $C_n$ is a normalization constant, $n \in \{1,2,\ldots \}$ labels the energy levels and $z_p= x\cos(p/2)$. We have introduced the dimensionless ratio $x=2h_x/h_z$. To get a normalized wave function in the infinite chain limit we
have set the coefficient in front of the increasing Neumann function solution $N_{l-\nu_{p,n}}(z_p)$ to zero.
The numbers $\nu_{p,n}$ are obtained by the other boundary condition, Eq.~(\ref{twodomainwalleq}) for $l=1$\cite{Fogedby78}: 
\be
     J_{-\nu_{p,n}}(z_p)=0.
\ee
For $z_p >\sim 1$, $\nu_{p,n}$ will depend on $p$ for the lowest values of $n$, thus the lowest lying spin-cluster excitations will be dispersive\cite{Shinkevich1}.  
For $z_p <\sim 1$,  $\nu_{p,n} \approx n$. Thus for $x<1$, $E_{p,n} \approx J_z + h_z n$ is independent of $p$ and the energy levels are equally spaced with a separation $h_z$. This is the WZL and we expect finite temperature neutron scattering to reveal transitions within the ladder with characteristic frequencies $\omega = k h_z$, where $k \in {\cal Z}$. The wave function describing the spin-cluster excitations in this WZL is a superposition of domains of different lengths $l$, each with weight $\psi_{p,n}(l) = C_n J_{l-n}(z_p)$ for $l>0$. 

\section{Dynamical structure factor}
We will now use the spin-cluster excitations to calculate the spectral weight of the low frequency peaks analytically.  
The transverse dynamical structure factor is 
\be
  S^{xx}(q,\omega)= \f{1}{Z} \sum_{i} \int_{-\infty}^\infty \! \! \f{dt}{2\pi} e^{i \omega t} \langle i | e^{-\beta H} S^x_{-q}(t) S^{x}_q(0) |i \rangle
 \label{SQWdef}
\ee
where $Z$ is the partition function and $\beta$ is the inverse temperature. The spin operators in momentum space
are $S^x_q = \sum_r e^{-iqr} S^x_r$ and have time dependence $S^x_{q}(t)=e^{i H t} S^x_{q} e^{-i H t}$. The sum is taken over all energy eigenstates of the system. The time integral can be performed by inserting an extra sum over energy eigenstates 
\be
  S^{xx}(q,\omega)= \f{1}{Z} \sum_{ij} \delta(\omega-(E_j-E_i)) e^{-\beta E_i} |\langle j| S^{x}_q |i \rangle|^2.
\ee
We will assume that the energy eigenstates can be written as states of almost independent spin-clusters. Thus we specify an energy eigenstate with $d$ spin-clusters as $|n_1,p_1,n_2,p_2,\ldots,n_d,p_d \rangle$ where $n_r$ labels the internal quantum number of spin-cluster $r$, and $p_r$ its momentum. we assume that this state has energy $E= J_z d+ h_z \sum_{r=1}^d n_r$.
Thus we neglect any interaction energies between different spin-clusters. Such many-spin-cluster states are orthogonal and complete for $x=0$. For finite $x$ we expect orthogonality and completeness to hold approximately when the density of spin-clusters is not too high.
 
The internal quantum number $n_r$ is an integer, so the dynamical structure factor will have peaks at low energies when $E_j-E_i = k h_z$ for $k \in {\cal Z}$. 
To get the spectral weight $S^{xx}_{k}(q)$ of the peak at $\omega=k h_z$ we integrate over frequencies in the vicinity of $k h_z$
\be
   S^{xx}_k(q) = \f{1}{Z} \sum_{ij} e^{-\beta E_i} | \langle j| S^x_q |i \rangle|^2 |_{E_j-E_i=kh_z} \label{SkqA3}
\ee
where the sum over states is restricted to states $|j\rangle$ which energies differ from $|i\rangle$ by $k h_z$.

When the operator $S^x_q$ acts on the state $|i \rangle$ it gives a sum over states $|i^\prime \rangle$ which each differ from state $|i \rangle$ by having a single spin overturned. Many of these states do not contribute in Eq.~(\ref{SkqA3}) as they have an almost zero overlap with states $|j \rangle$ having energy $E_j$ that differ from $E_i$ by $k h_z$. This is in particular true when $S^x$ creates(destroys) two domain-walls. This increases(decreases) the energy by $J_z$ and will therefore not contribute to low energy peaks at $\omega=k h_z$. The only states $|i^\prime \rangle$ that contribute are those which is the result of flipping a spin so as to displace an already existing domain-wall. When the spin-clusters are not too densely packed we can treat the spin-clusters independently. In that case the matrix element reduces to the matrix element between two single spin-cluster states as the other spin-clusters are unchanged. However, at the temperatures considered here, one also needs to go beyond the independent approximation and take into account that the movement of a domain-wall can cause it to collide with the neighboring spin-cluster. 
The result of such a collision is the disappearance of two domain-walls and a merger of two domains. The energy of such a final state differs by the initial state by roughly $J_z$, and should therefore not be counted as a contribution to the low frequency peaks.  We thus need to explicitly exclude final states $|i^\prime \rangle$ where a domain-wall is moved such as to merge two domains. We achieve that by including projections $\deltaLL$ which meaning is that the domain $r$ in state $|i \rangle$ must be such that its left domain-wall can be moved one lattice spacing to the left without touching another domain-wall for there to be a contribution.
We can therefore write the matrix element that contribute to peaks at $kh_z$ as
\be
 \sum_j  | \langle j | S^x_q | i \rangle |^2 = \sum_{r=1}^{d_i} | A_{n_rp_r} \deltaLL   + A_{n_rp_r}^* \deltaRR  + B_{n_rp_r}  + B_{n_rp_r}^*  |^2
\label{matrixelement}
\ee
where $d_i$ is the number of domains in state $|i \rangle$ and 
\begin{align}
A_{n_rp_r} &=\f{1}{2} \sum_{l=1}^\infty  \psi^*_{p_r+q,n_r+k}(l+1) \psi_{p_r,n_r}(l) e^{i(ql-p_r)/2}, \label{Aexpression} \\
B_{n_rp_r} &=\f{1}{2} \sum_{l=2}^\infty \psi^*_{p_r+q,n_r+k}(l-1) \psi_{p_r,n_r}(l) e^{i(ql+p_r)/2}, \label{Bexpression}
\end{align}
are matrix elements between two single-cluster states, one with quantum numbers $(n_r+k,p_r+q)$ and its left(right) domain-wall displaced one lattice spacing to the left compared to the other state which have quantum numbers $(n_r,p_r)$. The right movement is described by their complex conjugate. 

By multiplying out the right hand side of Eq.~(\ref{matrixelement}) we get 
\begin{align}
 S^{xx}_k(q) &= \f{1}{Z} \sum_i e^{-\beta E_i} \sum_{r=1}^{d_i} \left\{ \vphantom{\f{1}{2}}
\right. (\deltaLL+ \deltaRR) 
\nonumber \\
   &  \times \left[ |A_{n_rp_r}|^2  +(A_{n_rp_r}+A^*_{n_rp_r})(B_{n_rp_r}+B_{n_rp_r}^*) \right]
\nonumber \\
   & \left. + \deltaLL \deltaRR  \left[A_{n_rp_r}^2+A_{n_rp_r}^{*2} \right] 
+ (B_{n_rp_r}+B_{n_rp_r}^*)^2   
\vphantom{\f{1}{2}} \right\}
\end{align}
where we have set $\deltaLL \deltaLL = \deltaLL$ and $\deltaRR \deltaRR = \deltaRR$.

The first term, with $\deltaLL$,  is proportional to the probability to find a spin-cluster with quantum numbers $(n_1,p_1)$ whose left domain-wall can be expanded to the left without touching the neighboring domain (periodic boundary conditions is assumed). This probability does not depend on the spin-cluster index $r$, thus the sum over $r$ can be taken into account by replacing the probability with the number of occurrences. We approximate this quantity with the number of occurrences of $n$ consecutive down spins surrounded by one up spin to the right and {\em two} up spins to the left, $N_{\up \up n \up} \equiv \sum_i \langle U_{i-1} U_{i} D_{i+1} \cdots D_{i+n} U_{i+n+1} \rangle$,  where $U_i=1/2+S^z_i$ $(D_i=1/2-S^z_i)$ is the projection operator onto up(down) spins at site i. The brackets denote both quantal and thermodynamical average. The two up-spins to the left allow the displacement of the left domain-wall one lattice spacing to the left without touching a neighboring domain. For $x=0$ this approximation is exact. For small $x$ it is valid approximately as the wavefunction $\psi_{n,p}(l)$ is dominated by the term with $l=n$. Similarly the terms with $\deltaRR$ is approximated by $N_{\up n \up \up}$, $\deltaLL \deltaRR$ by $N_{\up \up n \up \up}$ and the terms without projections by $N_{\up n \up}$. 
This results in
\begin{align}
 S^{xx}_k(q) &= \sum_{n} \sum_{p} \left\{ \vphantom{\f{1}{2}}
    \left[ 2|A_{np}|^2 + (A_{np}+A_{np}^*)(B_{np}+B_{np}^*) \right] N_{\up \up n \up} \right. \nonumber \\
   & \left. + \left[ A_{np}^2+A_{np}^{*2} \right] N_{\up \up n \up \up} + (B_{np}+B_{np}^*)^2 N_{\up n \up}\vphantom{\f{1}{2}} \right\} \label{geneq}
\end{align}
where we have used inversion symmetry $N_{\up n \up \up}=N_{\up \up n \up}$.

At high temperatures, $T > h_x$, we can approximate the $N_{\up n \up}$s by setting $h_x=0$ and use transfer matrices\cite{Suzuki}. In the thermodynamic limit the number of $n$ down-spins surrounded by $m$ up-spins to the left and $l$ up-spins to the right is 
\be \label{Nupdownup}
 N_{\up \ldots \up \down \ldots \down \up \ldots \up} = N P_\up K_+^{m-1} W K_-^{n-1} W K_+^{l-1}
\ee
where $P_\up = 1/2 + \sinh(\beta h_z/2)[2(\lambda-\cosh(\beta h_z/2))]^{-1}$ is the probability of finding a certain spin to be up.    
$W=e^{-\beta J_z/2}/\lambda$ is the domain wall factor and $K_\pm=e^{\pm \beta h_z/2}/\lambda$ is associated with finding two adjacent spins that points up(+) (down(-)). The constant 
\be
\lambda = \cosh{(\beta h_z/2)}+ \left[ \sinh^2(\beta h_z/2)+e^{-\beta J_z} \right]^{1/2}
\ee
 is the largest eigenvalue of the transfer matrix.
Similarly one can find the number of domain-walls that separates exactly $n$ up-spins and $m$ down-spins. It is in the thermodynamic limit
\be \label{Nupdown}
  N_{\up \ldots \up \down \ldots \down}
= N \gamma K_+^{n-1} W K_-^{m-1}
\ee
where
$\gamma = e^{-\beta J_z/2}[2(\lambda-\cosh(\beta h_z/2))]^{-1}$. 

The expressions for $A_{np}$ and $B_{np}$, Eqs.~(\ref{Aexpression}) and (\ref{Bexpression}), involve an infinite sum over $l$. In order to evaluate this we add and subtract terms with negative values of $l$, see Appendix. Substituting the resulting expressions into Eq.~(\ref{geneq}) gives
\begin{align}
  S^{xx}_k(q) &= \f{1}{2} N_{\up \up \down} J_{1-k}(w_q) \left( J_{1-k}(w_q) -2J_{-1-k}(w_q) \right) \nonumber \\
&+ \f{1}{2} N_{\up \down} J_{-1-k}^2(w_q) \nonumber \\
&- \f{1}{2}\delta_{k,0} J_1^2(w_q) \sum_{n=1} \left( N_{\up n \up}  +2N_{\up \up n \up} + N_{\up \up n \up \up} \right) \cos(qn) \nonumber \\
&+ \Delta(q)  \label{answereq}
\end{align}
where $w_q = 2x \sin(q/2)$ and $\Delta(q) \sim {\cal O}(x^2)$ is a correction due to extending the sums to negative infinity
\begin{align}
\Delta(q) &=\delta_{k,0} \f{x^2}{32} \left( 3N_{\up \down \up} +8N_{\up \up \down \up} + N_{\up \up \down \up \up} \right) \nonumber \\
  &-\delta_{k,0} \f{w_q^2}{8} \left( 2N_{\up \down \up} + 6N_{\up \up \down \up}+ N_{\up \up \down \up \up} \right)  \nonumber \\
  &- \delta_{k,1} \f{x^2}{32} \left(7 N_{\up \up \down \up} + 2N_{\up \up \down \up \up} \right) \nonumber \\
  &+ \delta_{k,1} \f{w_q^2}{16} \left( 5N_{\up \up \down \up} + 2N_{\up \up \down \up \up} \right) + {\cal O}(x^4).
\end{align}

\section{Numerical results}
\begin{figure}[htb]
\begin{center}
    \includegraphics[clip,width=8.5cm]{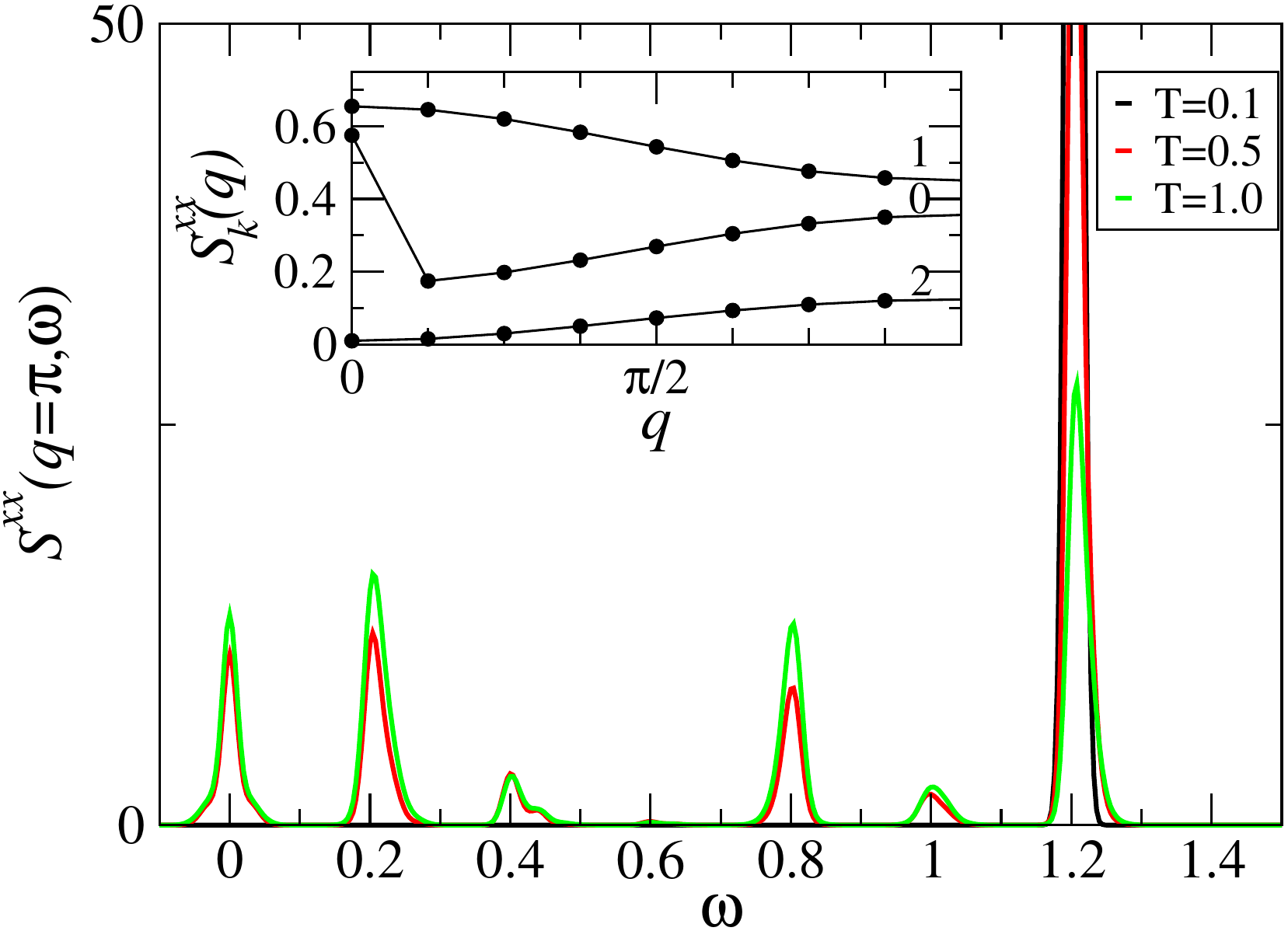} 
   \caption{(Color online) Dynamical structure factor $S^{xx}(q=\pi,\omega)$ obtained by exact diagonalization on a $16$-site chain for different temperatures indicated by the legends. $h_z=0.2$ and $h_x=0.1$. The delta functions have been broadened using Gaussians of width $\Delta=0.005$. 
The inset shows the integrated intensity of the peaks, $S^{xx}_k(q)$, vs $q$ for $k=1,0,2$ (top to bottom) at $T=0.5$.
   \label{Sqw}}
  \end{center}
\end{figure}

In order to validate Eq.(\ref{answereq}) we will calculate the dynamical structure factor $S^{xx}(q,\omega)$ numerically. To calculate $S^{xx}(q,\omega)$ at finite temperatures we have diagonalized the Hamiltonian (\ref{Hamiltonian}) of a chain of $N=16$ spins with periodic boundary conditions.  
Fig.~\ref{Sqw} shows the result as a function of frequency $\omega$ for a fixed value of momentum $q=\pi$ along the chain at different temperatures. The magnetic fields are $h_x=0.1$ and $h_z=0.2$. 

At low temperatures the only visible peak occurs at $\omega = J_z+h_z$. It can be attributed to the flipping of a single spin in an otherwise all spin-up background. The intensity of this spin-wave excitation diminishes as the temperature is increased, and new low-energy peaks appear at $\omega=kh_z$, where $k$ is an integer. This is the finite temperature signature of inter-level WZL transitions. At high temperatures the intensity of the $k=1$ peak becomes the same order of magnitude as the spin-wave peak. 

Some of the main features of the finite temperature dynamical structure factor in Fig.~\ref{Sqw} can be explained qualitatively using what one expects in the limit of vanishing $h_x$ when the Hamiltonian (\ref{Hamiltonian}) becomes purely classical. In that limit 
the peak at $\omega=h_z$ is associated with flipping a spin at a domain-wall. This process does not create new domain-walls, and the energy cost is therefore only the cost of one extra spin opposing the magnetic field. The peak at $\omega=J_z - h_z$ can be attributed to flipping a spin inside a domain of spins that are pointing opposite to the magnetic field. This creates two new domain-walls and makes one less spin pointing opposite to the field. These two processes can only occur at finite temperatures as they require the presence of domain-walls. Other features, such as the existence of peaks at $\omega=0,2h_z,J_z$ cannot be explained from considerations at $h_x=0$. 

When $h_x>0$ one expects generally that the location of the peaks in $S^{xx}(q,\omega)$ will acquire a dependence on the momentum $q$. This is true for the spin-wave peaks, but not for the low-energy peaks associated with the inter-level WZL transitions. They remain at frequencies $\omega=kh_z$ for all $q$. However, according to Eq.~(\ref{answereq}) their spectral weights should depend non-trivially on the momentum $q$. To estimate numerically the spectral weight of each low-energy peak at $\omega=k h_z$, $k=0,1,2$, we integrate the dynamical structure factor numerically over a small frequency interval chosen big enough to capture the full peak. We choose $\omega \in [ (k-1/2)h_z,(k+1/2)h_z ]$, and label the resulting integrated peak intensity $S^{xx}_k(q)$ as in the analytical calculation. A plot of these as functions of $q$ is shown in the inset of Fig.~\ref{Sqw} for $T=0.5$. The $k=1$ peak has the biggest spectral weight which decreases with increasing $q$, while the spectral weights of the $k=0,2$ peaks increase with $q$, except for the Bragg contribution at $\omega=0,q=0$ which is due to a finite ground state magnetization in the $x$-direction and is unrelated to BO. Our analysis is restricted to the lowest values of $k$ because at higher values of $k$ there is an overlap with the peak at $\omega=J_z-h_z$. To keep these separate we require $J_z-h_z >(k+1/2)h_z$.

\section{Comparisons}
To compare the analytical result Eq.~(\ref{answereq}) with our numerical results we consider each low-frequency peak (k) of the dynamical structure factor separately. 
\subsection{k=1}
\begin{figure}[h]
\begin{center}
    \includegraphics[clip,width=8cm]{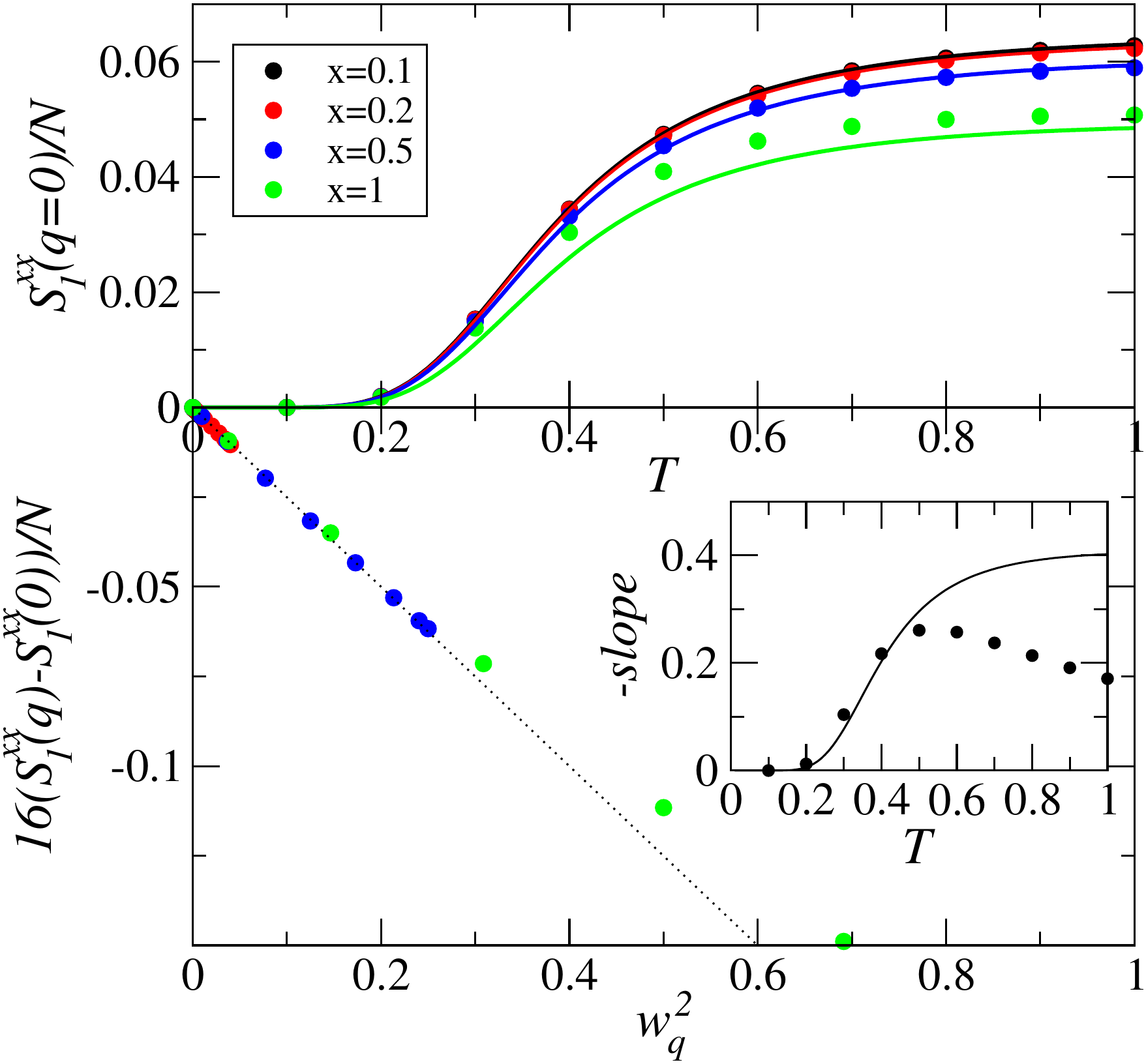} 
   \caption{(Color online) $k=1$ peak for $h_z=0.2$. The upper panel shows exact diagonalization results(symbols) for the $q=0$ intensity vs. T for different values of $x=2h_x/h_z$. The solid curves are gotten by Eq.~(\ref{k1result0}).  The lower panel shows how the $q$ dependence of the $k=1$ peak scales as $w_q^2$ for a fixed $T=0.5$. The inset shows the negative slope of such scaling curves vs. $T$ for $x=0.2$. The solid curve is the expression $(6N_{\up \up \down} - 5 N_{\up \up \down \up}- 2 N_{\up \up \down \up \up})/N$.  
   \label{k1figure}}
  \end{center}
\end{figure}

The expression for $S^{xx}_1(q)$ simplifies considerably at $q=0$,
\begin{align}
  S^{xx}_{k=1}(0) &= \f{1}{2} N_{\up \up \down} - \f{x^2}{32} \left( 7 N_{\up \up \down \up} + 2N_{\up \up \down \up \up} \right). \label{k1result0}
\end{align}
The upper panel of Fig.~\ref{k1figure} shows how the expression Eq.~(\ref{k1result0}) compares to the exact diagonalization results for $S^{xx}_1(q=0)$. It is rather accurate for $x<1$, but underestimates the numerical result for $x=1$.  
According to the analytical result the dynamical structure factor for $k=1$ depends on $q$ through the variable $w_q^2$. Expanding for small $w_q^2$ we find
\begin{align}
  S^{xx}_{k=1}(q) &= S^{xx}_1(0) - \f{w_q^2}{16}  \left( \vphantom{\f{1}{2}} 6N_{\up \up \down} -  5N_{\up \up \down \up} - 2N_{\up \up \down \up \up} \right) \nonumber \\
   & \qquad + {\cal O}(x^4) \label{k1result}
\end{align}
The lower panel of Fig.~\ref{k1figure} shows the exact diagonalization results for $S^{xx}_1(q)-S^{xx}_1(0)$ plotted as a function of $w_q^2$ for different values of $x$ and $q$ at the same temperature. The data points for $x<1$ fall on the same curve, and an asymptotic linear dependence on $w_q^2$, shown as the dotted line, is seen for small $w_q$. In the inset we have plotted the negative slope of this dotted line, the point at $T=0.5$. Repeating the scaling plot for several temperatures we get the remaining points. The solid curve shows the analytical result for the slopes, Eq.~(\ref{k1result}). For $T < 0.5$ the analytical results agree well with the numerical results, while at higher $T$ the analytical result overestimates the slope. The reason for this is probably that interactions between spin-clusters becomes dominant at high temperatures and densities and must be treated more accurately. 

\subsection{k=2}
The $q=0$ part of the $k=2$ peak is small and of $\order{x^4}$, whereas $S^{xx}_2(q)$ depends on $q$ through the variable $w_q^2$. The leading behavior for small $w_q^2$ is
\be \label{k2expression}
  S^{xx}_{k=2}(q) = S^{xx}_{2}(0)+N_{\up \up \down} \f{w_q^2}{8}+ {\cal O}(x^4).
\ee
\begin{figure}[h]
\begin{center}
    \includegraphics[clip,width=8cm]{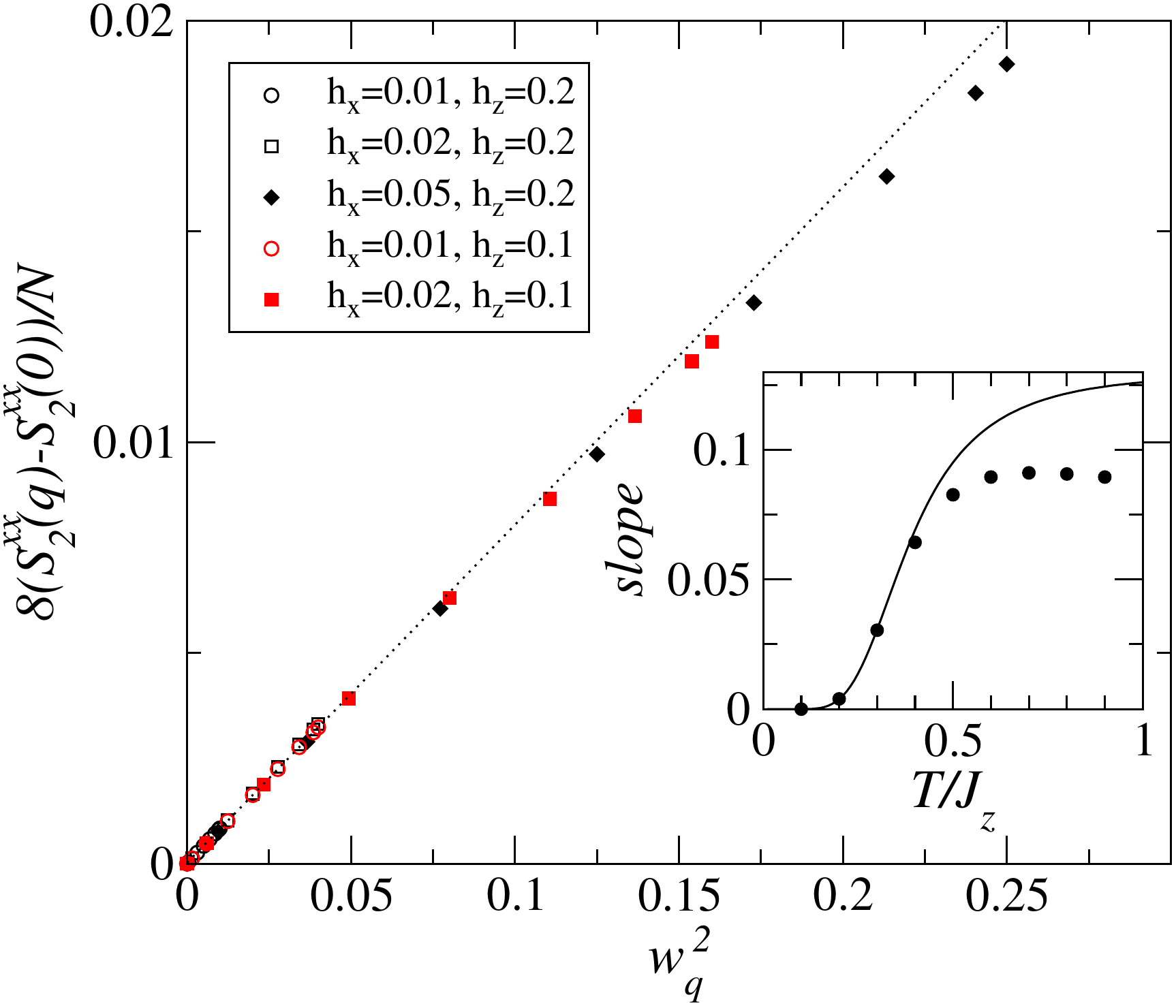} 
   \caption{(Color online) Scaling of the spectral weight of the $k=2$ peak with $w_q^2$. The main panel shows $8(S^{xx}_{2}(q)-S^{xx}_2(0))/N$ from exact diagonalizations vs. $w_q^2$ at $T=0.5$ for several values of magnetic fields indicated by the legends. The dotted line is the asymptotic linear behavior for small $w_q^2$. The inset shows the asymptotic slopes of the scaling curves for $h_x=0.02,h_z=0.2$ vs. $T$. For comparison $N_{\up \up \down}/N$ from Eq.~(\ref{Nupdown}) is plotted as the solid curve.    
   \label{k2figure}}
  \end{center}
\end{figure}   
In Fig.~\ref{k2figure} we have plotted $S^{xx}_2(q)-S^{xx}_2(0)$ from exact diagonalization vs. $w_q^2$ for several values of $q$ and magnetic fields at a temperature $T=0.5$. The points fall on the same curve demonstrating scaling with $w_q^2$. The dotted line shows the asymptotic linear behavior of the curve. It was obtained by lumping all the data points into one set, and fitting them to a quadratic function. In the inset we show the slope of this curve together with slopes of scaling functions obtained at other temperatures and compare them with Eqs.~(\ref{Nupdown}) and (\ref{k2expression}). As in the $k=1$ case the comparison is good up to $T \sim 0.5$.  

\subsection{k=0}
\begin{figure}[h]
\begin{center}
    \includegraphics[clip,width=8cm]{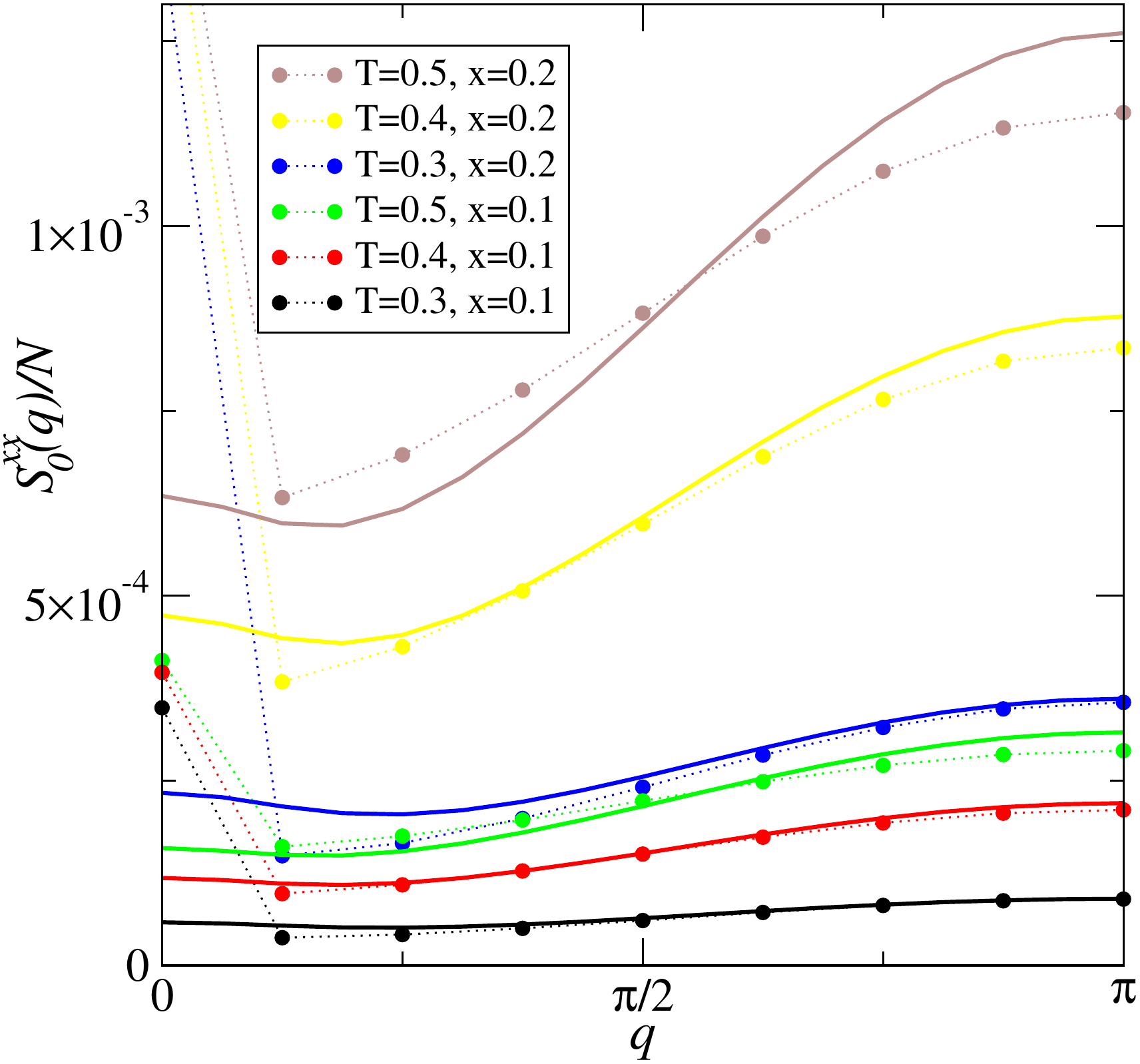} 
   \caption{(Color online) Spectral weight of the $k=0$ peak as functions of momentum for different values of temperature and magnetic field ratios $x$ as indicated by the legends. $h_z=0.2$. The solid dots are exact diagonalization results. The solid curves are the analytical result Eq.~(\ref{k0expression}).  
   \label{k0figure}}
  \end{center}
\end{figure}

For the zero energy peak, $k=0$, we need to evaluate the sum over $n$ in Eq.~(\ref{answereq}). Using the classical expressions, Eq.~(\ref{Nupdownup}), we get
\begin{align}
\sum_{n=1} &\left( N_{\up n \up} +2N_{\up \up n \up } + N_{\up \up n \up \up} \right) \cos(q n) \nonumber \\
&= \left( N_{\up \down \up} + 2N_{\up \up \down \up} + N_{\up \up \down \up \up} \right) \f{ \cos(q)-K_-}{1+K_-^2-2K_- \cos(q)}.
\end{align}
Expanding to order ${\cal O}(x^2)$ we find
\begin{align}
  S^{xx}_{0}(q) &= \f{x^2}{32} \left( 3N_{\up \down \up} +8N_{\up \up \down \up} + N_{\up \up \down \up \up} \right) \nonumber \\
&+\f{w_q^2}{8} \left( N_{\up \down} + 3 N_{\up \up \down} \right) \nonumber \\
&-\f{w_q^2}{8} \left( N_{\up \down \up} + 2N_{\up \up \down \up} + N_{\up \up \down \up \up} \right) \nonumber \\
&\qquad \times \f{ \cos(q)-K_-}{1+K_-^2-2K_- \cos(q)} \nonumber \\
  & -\f{w_q^2}{8}\left( 2N_{\up \down \up} + 6N_{\up \up \down \up}+ N_{\up \up \down \up \up} \right). \label{k0expression}
\end{align}
This expression is a smooth function of $q$. Thus there are zero energy excitations also at a finite $q$. In addition to this expression, at $q=0$, there is a term coming from the finite ground state magnetization in the $x$-direction, i.e. from intra ground state transitions which cause the discontinuous behavior at $q=0$ seen in the inset of Fig.~\ref{Sqw} for $k=0$. This additional term is proportional to the ground state transverse magnetization squared. Although it is the dominant term, we will ignore it here as it is not related to BO and only occurs at $q=0$ and $\omega=0$. Fig.~\ref{k0figure} shows a comparison of the numerical results with the expression Eq.~(\ref{k0expression}) for several values of the temperature and ratios of magnetic fields. The agreement is reasonable for low $x$-values and temperatures $T <\sim 0.5$.

\section{Cobalt niobate}
The material $\cobaltniobate$ is an experimental realization of a system of weakly coupled one-dimensional Ising-like ferromagnets.
It has been studied with neutron scattering and clear evidence of dispersive spin-cluster states have been observed in zero applied magnetic fields\cite{Coldea}. The fact that the excitations are dispersive at zero fields means that they cannot be interpreted as belonging to the WZL. Furthermore, it means that the Hamiltonian of $\cobaltniobate$ must contain extra terms in addition to the dominant Ising coupling\cite{Coldea}. A more detailed microscopic model which includes both internal magnetic fields and extra spin-spin couplings has been proposed for this material\cite{Kjall}. In particular, the large ratio of the internal fields $h^i_x/h^i_z \sim 10$ causes the spin-clusters to be dispersive over a large region of momentum space at low energies, and the WZL-ladder is present only near the zone boundary\cite{Rutkevich} which in the neutron scattering experiments is dominated by the ``kinetic bound state'' caused by the extra spin-spin couplings.   

\begin{figure}[h]
\begin{center}
    \includegraphics[clip,width=8.5cm]{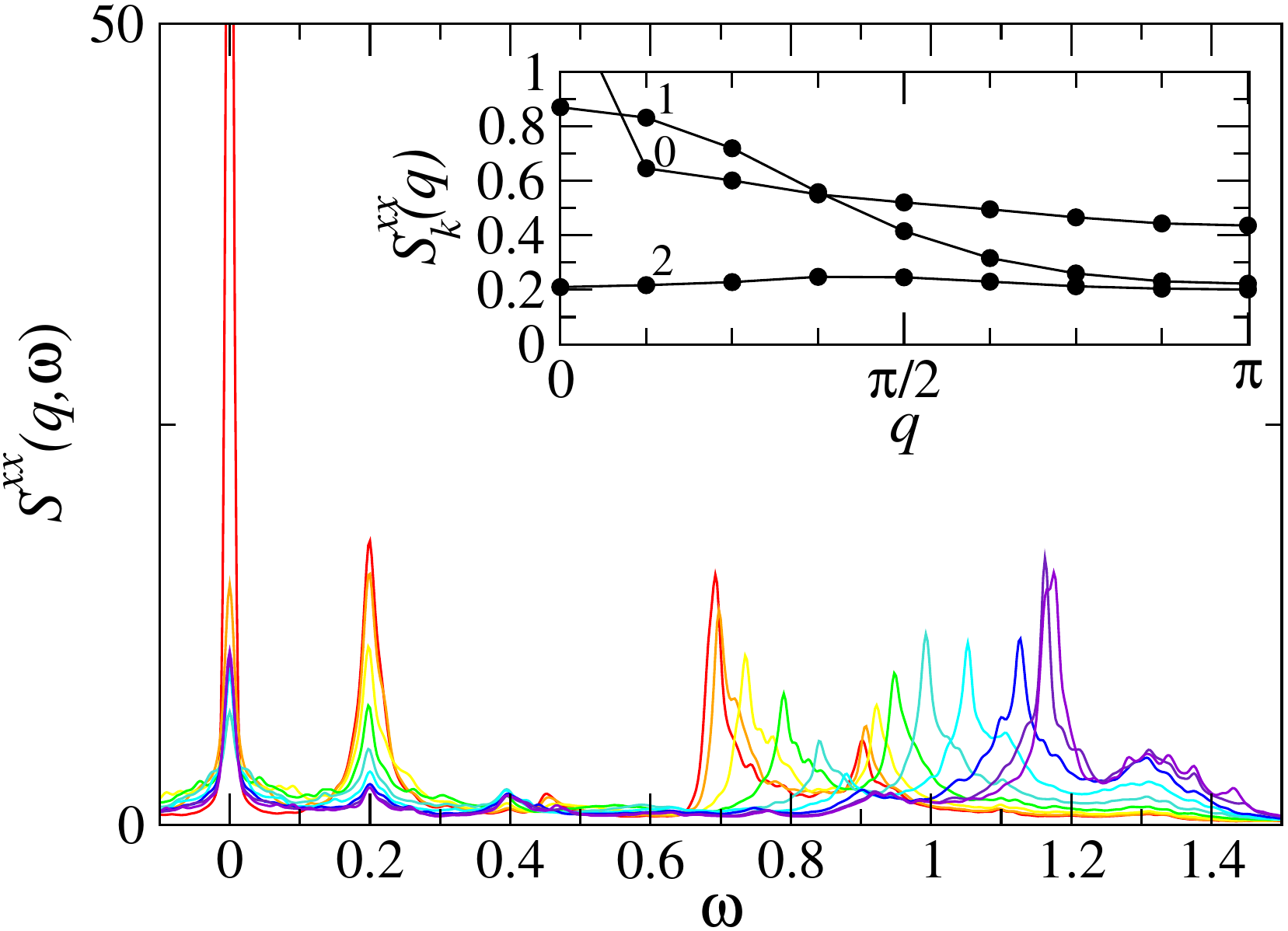} 
   \caption{(Color online) Exact diagonalization results on a 16-site chain for parameters relevant for the material $\cobaltniobate$. The main panel shows $S^{xx}(q,\omega)$ vs. $\omega$
. Each curve is for a specific $q$-value, and $q$ takes values from $0$(red) to $\pi$(violet). The magnetic fields are $h_x=0.1$, $h_z=0.2$ and the temperature is $T=0.5$. The curves are obtained by broadening the delta-functions using Gaussians of width $\Delta=0.005$. 
The inset shows the spectral weights $S^{xx}_k(q)$ for the low-frequency peaks with $k=1,0,2$ indicated by the labels.
   \label{Niobate}}
  \end{center}
\end{figure}

Still, one should expect to see the finite temperature signatures of MBO in $\cobaltniobate$ if the magnetic fields can be arranged so that $h_x/h_z <1$.  To investigate this closer we have repeated the exact diagonalization of the skew field Ising model including also additional terms\cite{Kjall}. The extra terms we have added to the Hamiltonian (\ref{Hamiltonian}) are $H_{xCoNb}=-J_{x} \sum_{i} \left( S^x_i S^x_{i+1} + S^y_i S^y_{i+1} \right) +J_{NN} \sum_i S^z_i S^z_{i+2}$ where $J_{x}=0.214$ and $J_{NN}=0.247$. The magnetic field components, including both internal and external magnetic field contributions, have been set to $h_x=0.1$, $h_z=0.2$, and the temperature is $T=0.5$. 
Fig.~\ref{Niobate} shows the resulting dynamical structure factor as a function of frequency for eight different momenta $q \in [0,\pi]$ plotted on top of each other. 

At high frequencies, $(\omega > \sim 0.6)$, there are dispersive peaks.  
For low frequencies one can clearly see the two non-dispersive peaks at $\omega=kh_z$, $k=0,1$ associated with transitions between WZL states even in the presence of the additional couplings. The spectral weights of these, and for a smaller peak at $k=2$, are shown in the inset as a function of momentum. The general trend is the same as without the extra couplings, except that the intensity of the $k=1$ peak shows a larger dependence on $q$ and the $k=0,2$ peaks have in general a larger intensity. 

\section{Conclusions}
We have investigated the finite temperature spectral signal of MBO in one-dimensional Ising models in a skew magnetic field using both analytical calculations and numerical exact diagonalizations. The main neutron scattering signature of MBO are finite temperature low-frequency peaks at $\omega=k h_z$, where the $k=1$ peak generally has the biggest spectral weight. While the peak frequency does not depend on momentum, its spectral weight does. We have shown that the spectral weight of the peaks can be calculated analytically when $h_x/h_z<1$ and $h_x < T <\sim J_z/2$ using an approach that treats the excitations of the chain as a gas of spin-cluster excitations. 

We have also investigated numerically how these results change when adding other couplings relevant for the material $\cobaltniobate$, and conclude that the extra couplings present will not destroy the MBO signatures.  


This research was supported by a grant NFR-213606 from the Research Council of Norway, and the numerical calculations were performed on the Abel computer cluster with a Notur grant nn4563k.

\begin{appendix}
\section{Bessel sums}
The expressions for the matrix elements needed in the calculation of the dynamical structure factor involve an infinite sum over $l$. Substituting in the Bessel function forms for $\psi$, Eq.~(\ref{psi}), for $x<1$, and omitting the normalization constant which is very close to unity we get 
\begin{align}
A_{np} &=\f{1}{2} \sum_{l=1}^\infty  J_{l+1-n-k}(z_{p+q}) J_{l-n}(z_p) e^{i(ql-p)/2}, \\
B_{np} &=\f{1}{2} \sum_{l=2}^\infty J_{l-1-n-k}(z_{p+q}) J_{l-n}(z_p) e^{i(ql+p)/2}.
\end{align}
In order to evaluate these sums we add and subtract terms with negative values of $l$. For $A_{np}$ we get
\begin{align}
A_{np} & =\f{1}{2} \sum_{l=-\infty}^\infty  J_{l+1-n-k}(z_{p+q}) J_{l-n}(z_p) e^{i(ql-p)/2} + \Delta A_{np} \nonumber \\
  & = \f{1}{2} J_{1-k}(w_q) e^{i(qn+\pi(1-k)-kp)/2} +\Delta A_{np}
\end{align}
where $w_q=2x \sin(q/2)$ and the sum has been evaluated using Graf's summation formula\cite{Watson}, and $\Delta A_{np}$ is correcting for the added terms with $l \leq 0$:
\begin{align}
\Delta A_{np} &= -\f{1}{2} \sum_{l=-\infty}^0 J_{l+1-n-k}(z_{p+q}) J_{l-n}(z_p) e^{i(ql-p)/2}  \nonumber \\
    & \approx \delta_{n,1} (-1)^k \f{z^k_{p+q} z_p}{2^{k+2} k!} e^{-ip/2}
\end{align}
where we have replaced the sum by its lowest order term in $x$ which is $\order{x^{2n+k-1}}$. 
In a similar way
\begin{align}
B_{np} &= -\f{1}{2} J_{-1-k}(w_q) e^{i(qn+\pi(1-k)-kp)/2} +\Delta B_{np}, \\
\Delta B_{np} &\approx \delta_{n,1} (-1)^k  \f{z_{p+q}^{1+k}}{2^{k+2} (k+1)!} e^{i(q+p)/2}.
\end{align}
\end{appendix}


\end{document}